\documentclass[prd,twocolumn,preprintnumbers,showpacs]{revtex4-1}
\usepackage{amsmath, mathrsfs, amssymb,amsfonts,amsthm,graphicx, epsf, dcolumn, yfonts}
\usepackage[hyperfootnotes=true]{hyperref}
\usepackage{subfigure}
\usepackage{color}
\usepackage{slashed}
\usepackage{setspace}
\usepackage{cancel}
\usepackage{wasysym}
\usepackage{hyperref}
\pdfoutput=1
\newcommand{\la}[1]{\label{#1}}
\newcommand{\eq}[1]{(\ref{#1})}
\newcommand{\prt}[1]{{\left( {#1} \right)}}
 \newcommand{\be}{\begin{equation}}
 \newcommand{\ee}{\end{equation}}
 \newcommand{\bea}{\begin{eqnarray}}
 \newcommand{\eea}{\end{eqnarray}}

\newcommand{\ff}{\frac}

\def\th{\theta}

\def\lab{\label}

\renewcommand*{\thefootnote}{\fnsymbol{footnote}}

\begin{document}

\title{Strongly-coupled anisotropic gauge theories and holography}
\author{Dimitrios Giataganas,$^\dag$ Umut G\"ursoy$^\ddag$ and Juan F. Pedraza$^\star$}
\affiliation{\vspace{1mm}$^\dag$Physics Division, National Center for Theoretical Sciences,
National Tsing-Hua University, Hsinchu 30013, Taiwan\\
$^\ddag$Institute for Theoretical Physics and Center for Extreme Matter and Emergent Phenomena,\\
Utrecht University, Leuvenlaan 4, 3584 CE Utrecht, The Netherlands\\
$^\star$Institute for Theoretical Physics, University of Amsterdam,
Science Park 904,\\ Postbus 94485, 1090 GL Amsterdam, The Netherlands}

\begin{abstract}\vspace{-2mm}
We initiate a non-perturbative study of anisotropic, non-conformal and confining gauge theories that are holographically realized in gravity by generic Einstein-Axion-Dilaton systems. In the vacuum our solutions describe RG flows from a conformal field theory in the UV to generic scaling solutions in the IR with generic hyperscaling violation and dynamical exponents $\theta$ and $z$. We formulate a generalization of the holographic c-theorem to the anisotropic case. At finite temperature, we discover that the anisotropic deformation reduces the confinement-deconfinement phase transition temperature suggesting a possible alternative explanation of inverse magnetic catalysis solely based on anisotropy. We also study transport and diffusion properties in anisotropic theories and observe in particular that the butterfly velocity that characterizes both diffusion and growth of chaos transverse to the anisotropic direction saturates a constant value in the IR which can exceed the bound given by the conformal value.

\end{abstract}

\renewcommand*{\thefootnote}{\arabic{footnote}}
\setcounter{footnote}{0}

\maketitle

\noindent \textbf{1. Introduction.}
\lab{sec::intro}
Quantum many body systems in three spatial dimensions with reduced rotational symmetry
have important realizations in Nature such as the quark-gluon plasma produced in non-central heavy ion collisions, or condensed matter systems described by anisotropic spin models e.g.~the anisotropic 3D Ising model. The rotational symmetry in such systems can be broken by application of an external source such as an electric or  magnetic field in one direction as in the various condensed matter experiments, by the geometry of the setting as in non-central heavy ion collisions, or by intrinsic properties of the interaction as in case of the anisotropic spin models or the Weyl semimetals \footnote{In this latter case isotropy is broken by separating the left and the right-handed Dirac cones along an axis in the momentum space.}.

Gauge-gravity duality \cite{adscft1} provides a natural avenue to study anisotropic QFTs in the presence of strong interactions. Most of the early gauge-gravity literature on anisotropic systems focuses either on scale-invariant systems or non-conformal but charged plasmas. Only the following three special cases have been studied: (i) initially conformal invariant systems where the isotropy and conformal symmetry is broken by the same mechanism, for example by a source that depends on a spatial direction as in \cite{Azeyanagi:2009pr, Mateos:2011tv,Mateos:2011ix,Giataganas:2012zy,Chernicoff:2012iq,Giataganas:2013hwa,Rougemont:2014efa, Fuini:2015hba,Arefeva:2016phb}. (ii) Lifshitz invariant systems with anisotropy as in \cite{Gursoy:2016tgf}. (iii) Non-conformal charged plasmas where the anisotropy is introduced by an external magnetic field in one spatial direction as in \cite{Rougemont:2015oea, Drwenski:2015sha, Gursoy:2016ofp}.

In this paper we initiate a study of uncharged,  non-conformal and anisotropic systems with strong interactions by means of the gauge-gravity duality. In particular, we consider a non-conformal, gapped  and {\em confining} 4D SU($N$) gauge theory  in the large-$N$ limit, obtained by deforming a strongly interacting fixed point in the UV by means of a scalar operator ${\cal O}$ with scaling dimension $\Delta$. We introduce the anisotropy by means of another operator $\tilde{\cal O}$ that we choose to be marginal, with a coupling that depends on one of the spatial coordinates. We then study influence of anisotropy on the RG flow at zero temperature and on the thermodynamic observables and transport coefficients at finite temperature. The gauge theory is realized in holography by the Einstein-Axion-Dilaton theory in 5 dimensions with a non-trivial potential for the dilaton. This potential can be chosen such that the vacuum state confines color and there exists a phase transition at finite $T_c$ above which a deconfined plasma state arises. We study this class of non-conformal, confining, anisotropic and strongly interacting theories holographically for a specific choice of the potentials, however, our qualitative results---which we discuss below---are independent of these choices and hold for the entire class.

The vacuum state of the theory is studied in section \ref{sec::scaling} and exhibits interesting qualitative features. In particular, we find that for marginal ${\cal O}$, that is $\Delta=4$, we find a non-trivial RG flow from the conformal fixed point in the UV to a Lifshitz-like hyperscaling violating theory in the IR with a range of possible dynamical and hyperscaling violating exponents $z$ and $\theta$ whose values determined by the choice of potentials in the dual gravitational theory\footnote{Generalizing examples in IIB supergravity \cite{Azeyanagi:2009pr,Mateos:2011tv} which only allows a fixed dynamical exponent $z=3/2$}. These models therefore open new ground for phenomenological applications in strongly interacting plasma physics.

An especially interesting question concerns how the confinement-deconfinement phase transition is affected by anisotropy. We study the phase diagram in section \ref{sec::thermo}
and discover that the anisotropic deformation decreases the confinement-deconfinement transition temperature. This is in accord with the recent lattice QCD studies \cite{Bali:2011qj,Bali:2011uf,Bali:2012zg,DElia:2012ems} that shows both chiral symmetry restoration and deconfinement occur at lower temperatures in the presence of an external magnetic field, a phenomenon coined  ``inverse magnetic catalysis'' (IMC). Note that magnetic field also breaks isotropy in a similar way as we do in our uncharged plasma. Yet, our finding brings a new twist in this story indicating that  IMC may occur even in uncharged plasmas.

Transport properties also exhibit surprising qualitative features. In particular, as we show in section 5, the butterfly velocity violates the ``universal bound'' conjectured in \cite{Mezei:2016wfz,Mezei:2016zxg}.

\noindent \textbf{2. Holographic setup.}
\lab{sec::setup}
The gravitational theory dual to our anisotropic field theory is defined by the Einstein-Axion-Dilaton action with generic functions $V$ and $Z$ that determine the potential energy for the dilaton field $\phi$ and its coupling to the axion field $\chi$:
\bea\label{action}
&&\quad\, S=\frac{1}{2\kappa^2}\int d^5x\,\sqrt{-g}\left[R+\mathcal{L}_M\right],\\
&&\mathcal{L}_M=-\frac{1}{2}(\partial \phi )^2+V(\phi )-\frac{1}{2}Z(\phi )(\partial \chi )^2,
\eea
where $\kappa^2\sim1/N^2$. Crucially, a linear axion ansatz automatically satisfies the equations of motion and breaks isotropy while preserving translation invariance:
\bea\label{metric}
\!\!\!\!\!ds^2&=&e^{2A(r)}\left[-f(r) dt^2+d\vec{x}_\perp^2 + e^{2h(r)} dx_3^2+\frac{dr^2}{f(r)}\right]\!,\\
\phi&=&\phi(r), \qquad \chi=a\, x_3.
\eea
The solution is asymptotically AdS near the boundary $r\to 0$ where $A\to -\log r$, $f\to 1$, $h\to 0$ and $\phi \to j\, r^{4-\Delta}$ \footnote{The asymptotic expansions for $\Delta=4$ are more subtle. For $\Delta=4$ we impose $\phi\to0$ in order to have an asymptotically AdS background.}. This solution generally corresponds to a non-conformal gauge theory whose IR dynamics dominated by the stress tensor $T_{\mu\nu}$ dual to the metric and a scalar operator $\mathcal{O}\sim\text{Tr}F^2$, similar to the scalar glueball operator in QCD (when it is marginal), here dual to the field $\phi$. We call the source of this operator $j$. The theory is in turn deformed by a space-dependent theta term $\tilde{\mathcal{O}}\sim \theta(x_3) \text{Tr}F\wedge F$ dual to the field $\chi$. The 5D Einstein-Axion-Dilaton theory can be realized in terms of D3/D7 branes in IIB string theory when $V=12$ and $Z=e^{2\phi}$ \cite{Azeyanagi:2009pr,Mateos:2011ix,Mateos:2011tv}. In this case the underlying field theory is conformal. We are however interested in non-conformal, in particular {\em confining} gauge theories that follow from a more generic choice of the potentials $V$ and $Z$ \cite{GK, GKN}.  A choice of the form \cite{Gubser:2008ny,Gubser:2008yx}
\be
V(\phi)=12\cosh(\sigma\phi)+b\,\phi^2,\qquad Z(\phi)=e^{2\gamma\phi},
\ee
with $b\equiv\frac{\Delta(4-\Delta)}{2}-6\,\sigma^2$, corresponds to a gauge theory with a scalar operator of scaling dimension $\Delta$ that confines color in the vacuum state for $\sigma \geq \sqrt{2/3}$ \cite{GKN}.

We observe that the holographic version of the c-theorem \cite{Freedman:1999gp} in QFT (or rather the ``a-theorem'' in 4D \cite{Komargodski:2011vj}) has a natural generalization in the anisotropic holographic theories. Introducing the domain-wall coordinate $du = \exp(A(r)) dr$ we find that
\be\lab{ctheo}
\frac{d}{du} \left\{ \left(\frac{dA}{du} + \frac13\frac{dh}{du}\right)e^{\frac{h}{3}}\right\} \leq 0 \, ,
\ee
which follows from Einstein's equations. Imposing the bulk null energy condition (NEC) recovers (\ref{ctheo}) but also leads to an additional monotonicity constraint,
\be \lab{ctheo2}
\frac{d}{du} \left( \frac{dh}{du}e^{h+4A}\right) \leq 0 \, ,
\ee
which can be used to define a second independent central charge for anisotropic theories (see also \cite{Hoyos:2010at,Liu:2012wf,Cremonini:2013ipa}). Both expressions inside the brackets of \eq{ctheo} and \eq{ctheo2} are monotonically decreasing and, while the first one reduces directly to $dA/du$ in the isotropic limit $h\to 0$, any linear combination between them may give the holographic analog of the a-function \cite{Freedman:1999gp}.

\noindent \textbf{3. Scaling solutions in the IR.}
\lab{sec::scaling}
The RG energy scale of the dual QFT in the ground state is determined by the scale factor $A$ of the metric (\ref{metric}) \cite{PolchinskiPeet}, which exhibits a non-trivial dependence on the  holographic coordinate $r$ when the potentials $V$ and $Z$ are not constant.   The IR region $r\to\infty$ corresponds to small values of $\exp(A)$ where the dilaton grows \footnote{The monotonicity is subject to a certain upper bound on $\sigma$ \cite{GJS} which we assume throughout the paper.} monotonically \cite{GK}. In this limit $V\sim 6\, e^{\sigma\phi}$ for $\sigma\geq0$.
We can derive the following scaling solutions in the IR limit:
 \bea\label{IRmetric}
\!\!\!\!\!ds^2&=&\tilde{L}^2 (ar)^{2\theta/3z}\left[\frac{-dt^2+d\vec{x}_\perp^2 + dr^2}{a^2r^2}+\frac{c_1~dx_3^2}{(ar)^{2/z}}\right]\!,\\
\lab{IRphi}
\phi&=&c_2\log(ar)+\phi_0.
\eea
Here $\tilde{L}$, $c_1$ and $c_2$ are constants depending on $z$ and $\theta$,
which are given in terms of $\gamma$ and $\sigma$ as
\be\la{z_th}
z=\frac{4 \gamma ^2-3 \sigma ^2+2}{2 \gamma  (2 \gamma -3 \sigma )},\qquad \theta=\frac{3 \sigma }{2 \gamma }.
\ee
These constants also depend on the free parameter $\phi_0$ which is set by the value of the source $j$.
For $\theta=0$ the solution exhibits a Lifshitz-like scaling
\be
\lab{scaling}
t\to \lambda t,\quad \vec{x}_\perp\to \lambda\vec{x}_\perp,\quad r\to \lambda r, \quad x_3\to\lambda^{\frac{1}{z}} x_3\, .
\ee
For $\theta\neq0$, the metric (\ref{IRmetric}) has the hyperscaling violation property and transforms covariantly under (\ref{scaling}) as
\be\lab{hyper}
ds\to\lambda^{\theta/3z}ds\,.
\ee
When the IR theory is connected to a heat bath, one obtains the finite temperature version of the scaling metric, which is now a black brane with blackening factor
\be
f(r)=1-\left(\frac{r}{r_H}\right)^{3+(1-\theta)/z}~,
\ee
where $r_H$ is the location of the horizon. The black brane metric is obtained by multiplying the $dt^2$ term by $f$ and the $dr^2$ term by $1/f$ in (\ref{IRmetric}). The entropy density of the plasma in the IR is obtained from the area of the horizon,
\be\label{IRent}
s=c_{\text{IR}}\,a^{-2-\frac{1-\theta}{z}}T^{2+\frac{1-\theta}{z}}/\kappa^2~,
\ee
where $c_{\text{IR}}$ is a constant and $T$ is the Hawking temperature
\be\la{temp1}
T=\frac{|3+(1-\th)/z|}{4\pi r_H}~.
\ee
Notice that the values $z$ and $\theta$
are constrained by the bulk null energy condition and the positivity of the specific heat $C_V = d\log s/d\log T$ as follows:
\bea\la{nec1}
&&(z-1)(1+3z-\theta)\geq0\,,\\\la{nec2}
&&\theta^2+3z(1-\theta)-3\geq0\,,\\
&&2z+1-\theta\geq0\,.\label{sh}
\eea
Combining these inequalities, we observe that for $z\ge 1$ the value of $\th$ is bounded from above $\th\le \th_{\rm{bound}}^{(-)}$ while for $z\leq0$ it is bounded from below $\th\ge \th_{\rm{bound}}^{(+)}$, with
\be \la{thbound}
 \th_{\rm{bound}}^{(\pm)}=\ff{1}{2} \prt{3 z \pm \sqrt{3}\sqrt{4 - 4 z + 3 z^2}}~.
\ee
The range $0<z<1$ is forbidden altogether. Thus, one derives interesting universal bounds on the IR scaling behavior of strongly interacting anisotropic plasmas from holography.
\begin{figure}[t!]
\includegraphics[angle=0,width=0.4\textwidth]{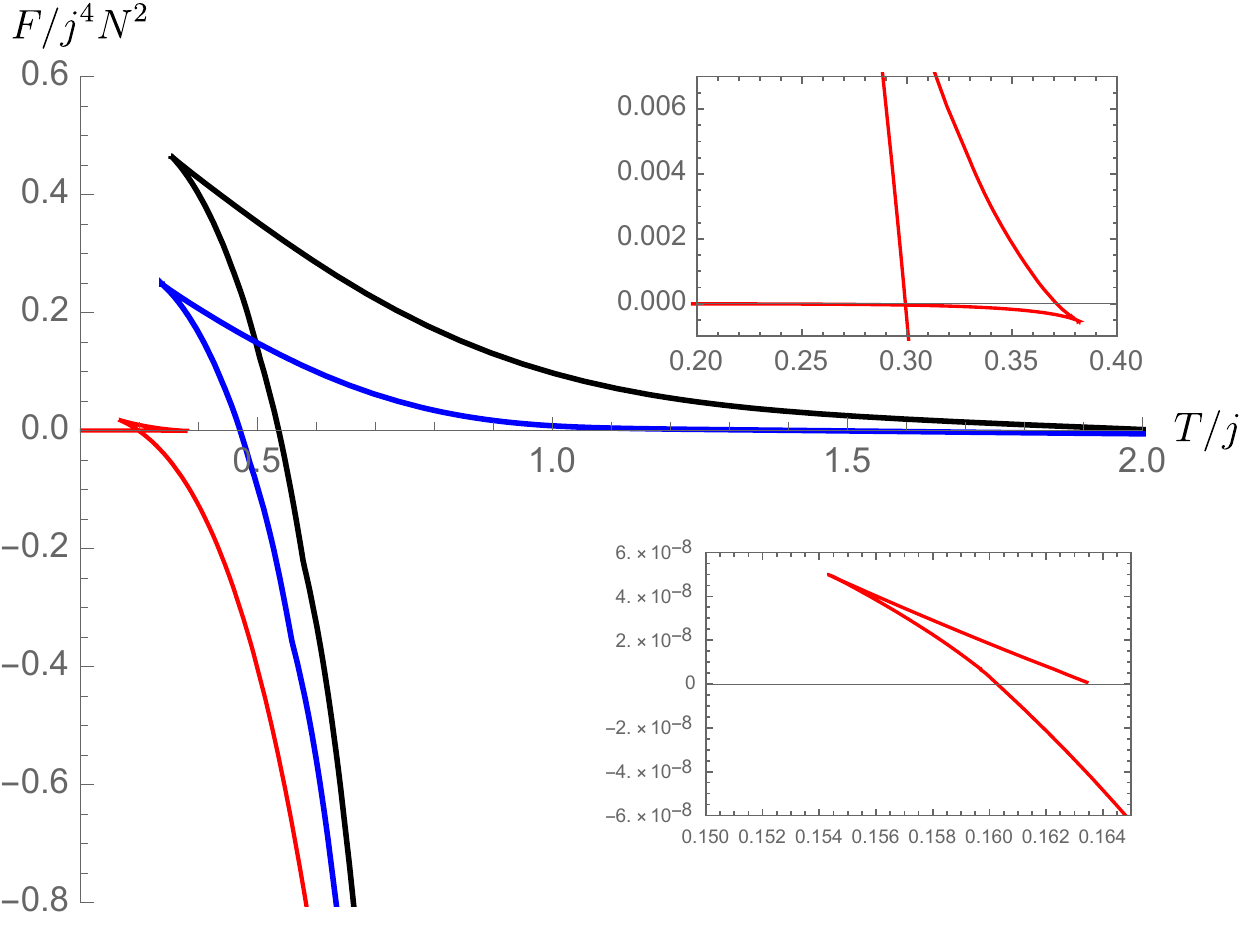}
\caption{\small Free energy as a function of $T$  for different values of the anisotropy parameter $a/j=0,1,3$ (black, blue and red curves respectively). The parameters $\sigma= \sqrt{2/3} + 1/10$, $\gamma = 1/5$, $\Delta = 3$ were chosen such that the undeformed theory is confining. The horizontal axis corresponds to the confined state while all other branches correspond to deconfined phases. The insets show details of an additional phase transition for large $a$, as discussed in the text.
\label{fig::free}}
\end{figure}

\noindent \textbf{4. Thermodynamics.}
\lab{sec::thermo}
Questions pertaining thermal equilibrium are answered by working out the free energy in the canonical ensemble, which, in the holographic description equals the Euclidean gravitational action (\ref{action}) appended by the Gibbons-Hawking and counterterm actions, evaluated on-shell. The counterterms in a generic Einstein-Axion-Dilaton theory were worked out in detail in \cite{Papadimitriou:2011qb} whose results we use but do not show here. Alternatively, one can calculate the background subtracted free energy directly by integrating the first law of thermodynamics $dF = -s dT$ for $j$ and $a$ held fixed \footnote{The physical coupling on the boundary $a\, x_3$ is kept fixed.}.

\begin{figure}[t!]
\includegraphics[width=0.4\textwidth]{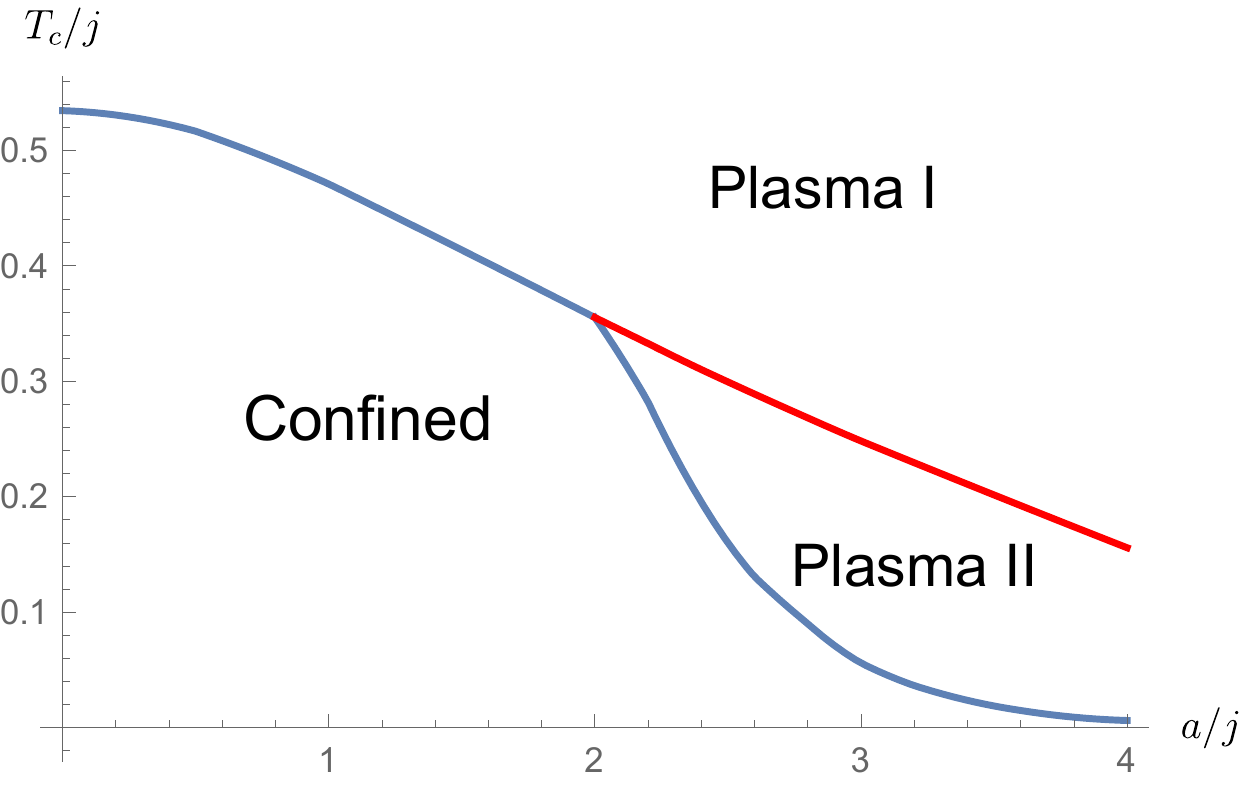}
\caption{\small Phase diagram of the system in the $a-T$ plane. We observe two phases, confined and ``plasma I", for $a/j<2.08$. For larger $a/j$ there exist three phases, confined, ``plasma I" and ``plasma II". The blue and red curves indicate lines of first order transitions.
\label{fig::tca}}
\end{figure}

In figure \ref{fig::free} we plot numerical results for the free energy as a function of $T$ for a particular confining theory. We will divide the analysis in two cases, small $a/j$ and large $a/j$. For small $a$ up to $a/j\approx 2.08$ there are three competing phases. First, there is the {\em confining ground state} heated up to temperature $T$.
The corresponding gravitational background is obtained from the black brane solution (\ref{metric}) by sending the mass to zero. This is the so-called thermal gas solution and is our reference background for the free energy computation. More specifically, the free energy of this phase is ${\cal O}(N^0)$, therefore it corresponds to the horizontal axis $F=0$. Second, we observe two phases of free energy ${\cal O}(N^2)$. These are the {\em deconfined, plasma phases} corresponding to black brane solutions (\ref{metric}) with a non-trivial blackening factor. One of these solutions, the ``small black brane'' (upper branches in figure  \ref{fig::free} for $a/j = 0$, 1) is always subdominant in the ensemble and can be ignored. Moreover this phase is thermodynamically unstable since $C_V\propto - d^2F/dT^2<0$, as can be read from the figure. The ``big black brane'' solution  (lower branches in figure \ref{fig::free} for $a/j = 0$, 1) dominates the ensemble for $T>T_c$. $T_c$ here is given by the point where the curves cross $F=0$. Therefore the system is in the deconfined phase above the critical temperature  $T_c$. This plasma phase is denoted as ``plasma I'' in figure \ref{fig::tca}. Below $T_c$ the system is in the confined phase. This phase transition is of confinement/deconfinement type and it is of first order. All of this is in accordance with improved holographic QCD \cite{PRL,longThermo}.

For $a/j > 2.08$ the phase structure becomes more complicated. As shown in figure \ref{fig::free} for the choice $a/j = 3$, there exists now four different black brane branches with free energy ${\cal O}(N^2)$ instead of the aforementioned two solutions, the small and the big black branes for $a/j< 2.08$. It is apparent from figure  \ref{fig::free} that two of them have positive specific heat, analog of the ``big black brane'' solution in the small $a$ case. These two solutions are denoted as ``plasma I'' and ``plasma II" in figure \ref{fig::tca}. There are two more black brane solutions analog of the ``small black brane'' solution in the small $a$ case. However, these are always subdominant and thermodynamically unstable, hence we can ignore them. As shown in figure \ref{fig::tca}, there are now two phase transitions. There is the confinement/deconfinement type first order transition, analogous to the small $a$ case, and there is a new first order transition between the two plasma phases at a higher critical temperature. Moreover, all of these dominant phases meet at a triple point at $a\approx 2.08$, $T/j\approx0.36$.

We observe that the confinement/deconfinement transition temperature $T_c$ decreases with increasing anisotropy $a$ as shown in figure \ref{fig::tca}. This is interesting in the context of inverse magnetic catalysis \cite{Bali:2011qj,Bali:2011uf,Bali:2012zg,DElia:2012ems}. It has been observed that the chiral symmetry breaking temperature decreases with increasing degree of anisotropy, induced by an external magnetic field $B$. Our finding suggests an alternative mechanism based only on anisotropy, as explained in the Discussion.

\noindent \textbf{5. Transport and Diffusion.}
\label{sec::transport}
\begin{figure}[t!]
\includegraphics[width=0.4\textwidth]{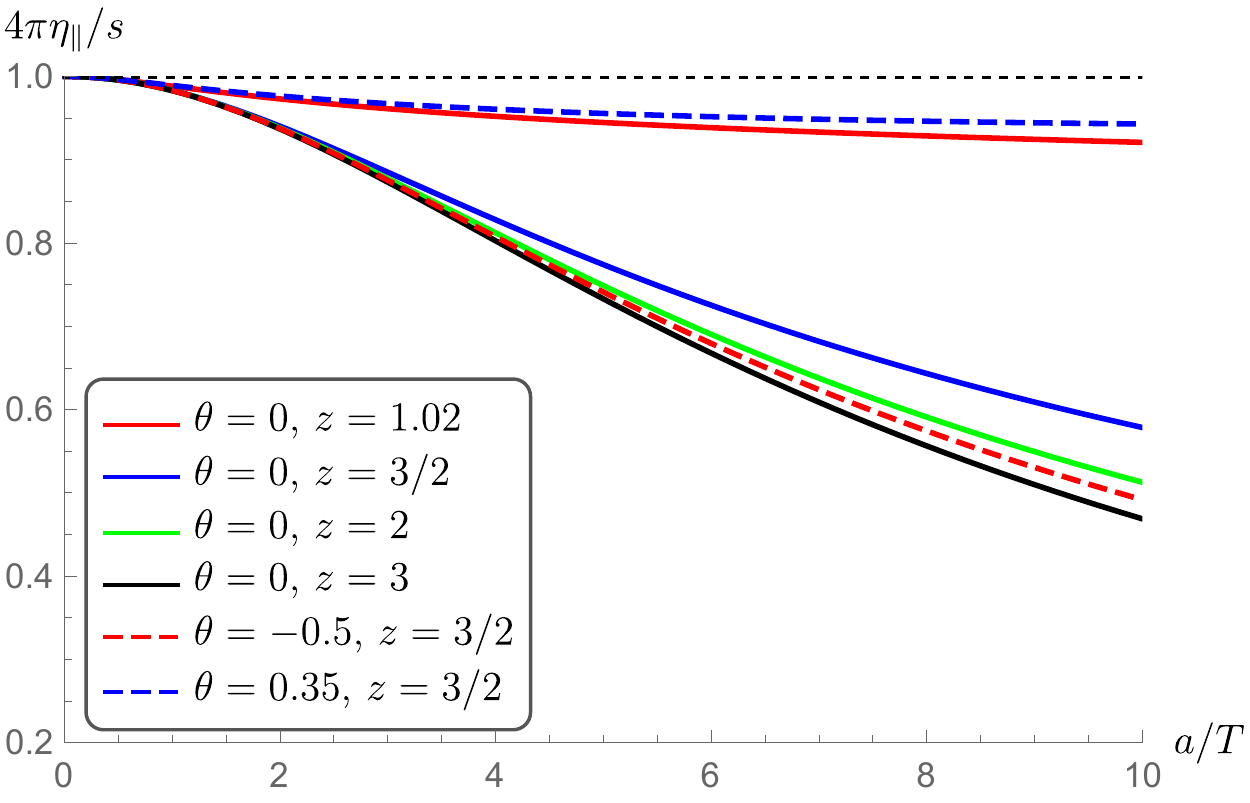}
\caption{\small The viscosity over entropy density ratio for several values of $\th$ and $z$. Increase of the scaling parameter $z$ leads to lower values of the ratio (solid lines), as well as a decrease of the value of $\th$ (dashed lines).
\label{figure:ratioall}}
\end{figure}
Holography is instrumental in the study of dissipative properties of the quark-gluon plasma as exemplified by the agreement between the holographic value of the shear viscosity $\eta/s = 1/4\pi$ \cite{Policastro:2001yc,Kovtun:2003wp} and experiment \cite{CasalderreySolana:2011us}. This universal value is violated in anisotropic systems \cite{Rebhan:2011vd,Mamo:2012sy,Jain:2015txa,Jain:2014vka,Erdmenger:2010xm} for the shear component parallel to the anisotropic direction, $\eta_{\parallel}$, while the component transverse to the anisotropic direction $\eta_\perp$ remains universal. A calculation shows that
$\eta_\parallel$ can be obtained from the near-horizon form of the metric (\ref{metric}),
\be\label{etaovers}
\ff{\eta_\parallel}{s}=\ff{1}{4\pi}\ff{g_{\perp\perp}}{g_{33}}\Bigg|_{r=r_h}~.
\ee
In figure \ref{figure:ratioall} we show the behavior of this quantity for the case $\Delta=4$ and $j=0$.
The curves are parametrized according to the properties of the IR geometry, i.e. the scaling exponents $z$ and $\theta$. We observe that the shear viscosity in the anisotropic direction is generally below the universal value $\eta/s=1/4\pi$, which is attained only in the UV.
In the IR limit one obtains
\bea\la{ratioT}
\ff{\eta_\parallel}{s}=\chi_{\text{IR}} \prt{\ff{a}{{T}}}^{\ff{2}{z}-2}\,.
\eea
The constant $\chi_{\text{IR}}$ depends generically on $z$, $\theta$ and $\phi_0$ (or equivalently $j$). The power of $a/T$ is independent of the hyperscaling violation exponent $\th$. However, $\th$ determines the allowed range of $z$ through \eq{nec1}-\eq{sh}; e.g. for $\th=0$ the exponent is in the range $(-2,0]$. More generally, the bounds imply that the power is always negative (for $z\neq1$) so $\eta_\parallel$ tipically vanishes in the deep IR.

Another interesting phenomenon is momentum diffusion, which is related to dissipation through shear via an Einstein relation. In holographic theories, diffusion is characterized by the time scale $\tau_L \sim 1/T$ and the ``butterfly velocity'' $v_B$ \cite{Blake:2016wvh}, both entering in the diffusion constant as $D \sim \hbar v_B^2/k_B T$. These quantities can be computed holographically through the near horizon dynamics. They also control the chaotic growth of the commutator $\langle [ W (t,x), V (0,0) ]^2\rangle \sim \exp[(t - x/ v_B)/\tau_L]$ for arbitrary hermitian operators $W$ and $V$, whose properties have been studied extensively recently in \cite{Hartnoll:2014lpa,Roberts:2014isa,Shenker:2014cwa,Blake:2016wvh,Roberts:2016wdl,Hartman:2017hhp}. Interestingly, the butterfly velocity provides a natural notion of a ``light cone'' even for non-relativistic systems, e.g. in \cite{Roberts:2016wdl} it was argued that $v_B$
acts as a low-energy Lieb-Robinson velocity.

\begin{figure}[t!]
\includegraphics[angle=0,width=0.4\textwidth]{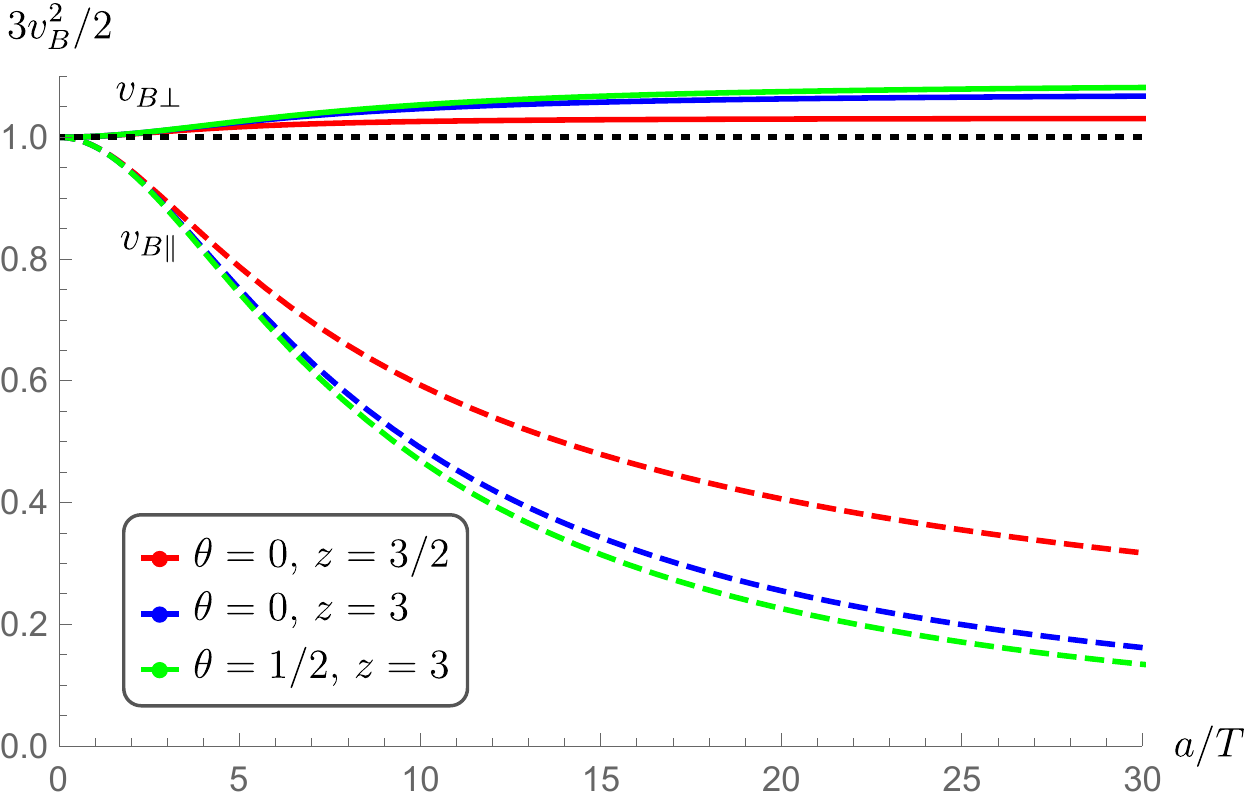}
\caption{\small Butterfly velocities $v_{B\perp}$ (solid lines) and $v_{B\parallel}$ (dashed lines). In the longitudinal direction the information diffuses slower with increasing anisotropy, with vanishing velocity in the IR. Perturbations in the transverse plane can propagate at a faster rate, with a new upper bound attained in the IR.
\label{figure:vball}}
\end{figure}

In anisotropic theories, there are two notions of butterfly velocities, $v_{B\parallel}$ and $v_{B\perp}$ \cite{Blake:2016wvh}, corresponding to the parallel and transverse directions, respectively \footnote{See \cite{Wu:2017exh,Jahnke:2017iwi} for related works.}. These can be obtained by studying the backreaction of an excitation at $\vec{x}=0$ sent from the asymptotic boundary into the bulk. The excitation solves the Poisson equation on the curved geometry with a delta-function source $\delta(\vec{x})$ and with an effective mass that corresponds to the screening length $\mu_\parallel$ or $\mu_\perp$ in the corresponding plasma. On the background (\ref{metric}) one finds,
\begin{equation}
\mu_\perp^2  =  \frac{f' (3A' + h')}{2}\bigg|_{r=r_h},\quad \mu_\parallel^2 = \mu_\perp^2 e^{2 h(r_h)}.
\end{equation}
The corresponding butterfly velocities $v_{Bi}^2=(2\pi T)^2/\mu_i^2$ approach the conformal value $v_B^2=2/3$ in the UV,
while in the IR one obtains
\bea
&&v_{B\perp}^2=\frac{1-\theta+3 z}{2(1-\theta+2z)}\,,\qquad v_{B\parallel}^2=\zeta_{\text{IR}}\left(\frac{a}{T}\right)^{\frac{2}{z}-2}\,,
\eea
where $\zeta_{\text{IR}}=4\pi v_{B\perp}^2\chi_{\text{IR}}$. Remarkably, $v_{B\perp}^2$ saturates to a universal value independent of $j$. These expressions are to be contrasted with the butterfly velocities for isotropic theories with hyperscaling violation \cite{Blake:2016wvh,Roberts:2016wdl}. In these works it was found that they scale generically as $v_{B}^2\sim(T_0/T)^{\frac{2}{z}-2}$, where $T_0$ is a UV scale, so $v_{B}^2\to0$ in the deep IR. In contrast, we find that $v_{B\perp}^2$ saturates to a constant for $a\gg T$. In the allowed range of $\theta$ and $z$, both $v_{B\perp}^2$ and $v_{B\parallel}^2$ are positive and smaller than 1 but, surprisingly, the value of $v_{B\perp}^2$ in the IR can exceed the conformal value $v_B^2=2/3$. For example, for $\theta=0$ and $z>1$, it always exceeds 2/3 and saturates {\em another bound} $v_{B\perp}^2\to 3/4$ as $z\to \infty$. In figure \ref{figure:vball} we plot $v_{B\perp}^2$ and $v_{B\parallel}^2$ as a function of $a/T$ for some generic examples.

\noindent \textbf{6. Discussion.}
\lab{sec::discussion}
In this work we found several qualitative features of strongly coupled anisotropic systems both in the ground state and at finite temperature. Studying confining plasmas, we found that the confinement-deconfinement phase transition temperature decreases with anisotropy. Decrease in $T_c(a)$ with $a$ resembles the phenomenon of inverse magnetic catalysis where both the confinement-deconfinement and the chiral phase transition temperatures decrease with the magnetic field $B$. The main difference is that there are no charged fermionic degrees of freedom in our case; our plasma is neutral. This suggests an explanation for IMC, alternative to e.g. \cite{SzaboBruckmann,MiranskyShovkovy}: that, anisotropy---which is caused by $B$ instead of $a$ in those examples---could be responsible for the phenomenon instead of charge dynamics. It is tempting to conjecture this as a generic property in a large class of confining anisotropic theories at strong coupling. In fact, studies with different holographic constructions \cite{upcoming} showing a destructive effect of anisotropy on the chiral condensate strongly support our claim. Whether there is a direct field theory argument in support of this conjecture remains to be seen. Nonetheless, it is interesting to note that the fact that anisotropy itself may be responsible for inverse magnetic catalysis could in principle be checked by (anisotropic) lattice calculations!

The ground state in our theories are generically characterized in the IR, by dynamical and hyperscaling violating exponents $z$ and $\theta$. Theories are thermodynamically stable for a wide range of these exponents. We also obtained a generalization of the holographic c-theorem valid for anisotropic theories in equation (\ref{ctheo}). It would be interesting to work out the implications of this for RG flows, and to provide a proof directly in quantum field theory.

Finally, as expected from previous works, we found $\eta_\parallel/s$ to be generically smaller than the so-called universal result $1/4\pi$. We also found that $v_{B\perp}^2$ can exceed the bound conjectured in \cite{Mezei:2016wfz,Mezei:2016zxg}. There is no contradiction, since these papers assumed isotropy. However, this result is surprising: it implies a new upper bound for the transfer of quantum information, enhanced here by the effects of anisotropy. It will be interesting to figure out the field theoretic reason behind these observations. It will also be interesting to find realization of our findings in physical systems. In connection to this we refer to \cite{Samanta:2016pic,Samanta:2016lsh} as way to measure the anisotropic shear viscosity of a strongly interacting, ultra-cold, unitary Fermi gas confined in a harmonic trap.

\begin{acknowledgments}
The authors acknowledge useful conversations with Mariano Chernicoff, Chong-Sun Chu, Viktor Jahnke, David Mateos and Phil Szepietowski. This work is partially supported by the Ministry of Science and Technology of Taiwan under the grants 101-2112-M-007-021-MY3 and 104-2112-M-007 -001 -MY3, the Netherlands Organisation for Scientific Research (NWO) under the VIDI grant 680-47-518 and a VENI grant, and the Delta-Institute for Theoretical Physics ($\Delta$-ITP), which is funded by the Dutch Ministry of Education, Culture and Science (OCW).
\end{acknowledgments}

\bibliography{botany}

\end{document}